\documentclass[twocolumn,trackchanges]{aastex7}

\usepackage{amsmath,xcolor}
\usepackage{float}
\usepackage{array}

\begin{document}

\title{Irradiated Atmosphere V: Effects of  Vertical-Mixing induced Energy Transport on the Inhomogeneity}

\correspondingauthor{Dong-dong Ni}
\email{ddni@nju.edu.cn}

\correspondingauthor{Cong Yu}
\email{yucong@mail.sysu.edu.cn}

\author[0000-0002-0447-7207]{Wei Zhong}
\affiliation{Institute of Science and Technology for Deep Space Exploration, Nanjing University, Suzhou, 215163, People's Republic of China}
\email{}

\author[0009-0004-1986-2185]{Zhen-Tai Zhang}
\affiliation{School of Physics and Astronomy, Sun Yat-Sen University, Zhuhai, 519082, People's Republic of China}
\affiliation{CSST Science Center for the Guangdong-Hong Kong-Macau Greater Bay Area, Zhuhai, 519082, People's Republic of China}
\affiliation{State Key Laboratory of Lunar and Planetary Sciences, Macau University of Science and Technology, Macau, People's Republic of China}
\email{}

\author[0000-0002-0378-2023]{Bo Ma}
 \affiliation{School of Physics and Astronomy, Sun Yat-Sen University, Zhuhai, 519082, People's Republic of China}
\affiliation{CSST Science Center for the Guangdong-Hong Kong-Macau Greater Bay Area, Zhuhai, 519082, People's Republic of China}
\email{}

\author[0000-0003-2278-6932]{Xianyu Tan}
\affiliation{Tsung-Dao Lee Institute \& School of Physics and Astronomy, Shanghai Jiao Tong University, Shanghai 201210, China}
\email{}

\author[0000-0002-0483-0445]{Dong-dong Ni}
\affiliation{Institute of Science and Technology for Deep Space Exploration, Nanjing University, Suzhou, 215163, People's Republic of China}
\affiliation{State Key Laboratory of Lunar and Planetary Sciences, Macau University of Science and Technology, Macau, People's Republic of China}
\email{}

\author[0000-0003-0454-7890]{Cong Yu}
\affiliation{School of Physics and Astronomy, Sun Yat-Sen University, Zhuhai, 519082, People's Republic of China}
\affiliation{CSST Science Center for the Guangdong-Hong Kong-Macau Greater Bay Area, Zhuhai, 519082, People's Republic of China}
\affiliation{State Key Laboratory of Lunar and Planetary Sciences, Macau University of Science and Technology, Macau, People's Republic of China}
\affiliation{International Centre of Supernovae, Yunnan Key Laboratory, Kunming 650216, People's Republic of China}
\email{}

\begin{abstract}

Atmospheric variations over time and space boost planetary cooling, as outgoing internal flux responds to stellar radiation and opacity. Vertical mixing regulates this cooling. Our study examines how gravity waves or large-scale induced mixing interact with radiation transfer, affecting temperature inhomogeneity and internal flux. Through the radiative-convective-mixing equilibrium, mixing increases temperature inhomogeneity in the middle and lower atmospheres, redistributing internal flux. Stronger stellar radiation and mixing significantly reduce outgoing flux, slowing cooling.  With constant infrared (IR) opacity, lower visible opacity and stronger mixing significantly reduce outgoing flux.  Jensen's inequality implies that greater spatial disparities in stellar flux and opacity elevate the ratio of the average internal flux in inhomogeneous columns relative to that in homogeneous columns. This effect, particularly pronounced under high opacity contrasts, amplifies deep-layer temperature inhomogeneity and may enhance cooling. However, with mixing, overall cooling is weaker than without, as both the averaged internal flux of the inhomogeneous columns and that of the homogeneous column decline more sharply for the latter. Thus, while vertical mixing–induced inhomogeneity can enhance cooling, the overall cooling effect remains weaker than in the non-mixing case. Therefore, vertical mixing, by regulating atmospheric structure and flux, is key to understanding planetary cooling.

\end{abstract}


\section{Introduction} 

Planetary surfaces and atmospheres are rarely uniform; instead, they exhibit spatial and temporal inhomogeneity \citep{ZhangXi2023ApJ}. Here, inhomogeneity refers to the non-uniform distribution and variability of energy input, material properties, and dynamical processes across different locations and times. In practice, this means that key properties such as temperature, albedo \citep{2023Chen}, and chemical composition are not constant but vary over planetary scales. For instance, tidally locked exoplanets maintain permanent day–night temperature contrasts \citep{2002A&AGuillot}, while patchy cloud structures evolve dynamically \citep{2013A&AParmentier,2017MNRAS...Hubeny,2019ApJPowell,2023Natur.Feinstein}.

Planetary inhomogeneity arises from several factors. Uneven stellar flux results from the planet’s spherical shape, orbital eccentricity, and axial tilt, while rotation drives diurnal cycles that modulate energy input \citep{2021MNRAS.Tan,2021MNRAS.Tan502,2023ApJ...Zhang...Inhomogeneity2}. Variations in surface properties, including thermal inertia, albedo, emissivity, and composition (e.g., water and ice), further contribute to spatial heterogeneity. The atmospheric circulation redistributes heat and chemical species  \citep{2016ApJ...816...96H,2016ApJ...829..104H}, while photochemical processes create localized tracers. 
In addition to these effects, stellar irradiation itself is often distributed unevenly across a planet, and atmospheric opacity can vary significantly with location due to clouds, composition, and dynamics. These spatial contrasts in stellar flux and opacity represent two key forms of planetary inhomogeneity.

Together, these processes alter radiative fluxes and opacity, influence energy transport and climate regulation \citep{1988Kasting,2023ApJ...Mukherjee,2023Innes}, modify chemical distributions \citep{2023A&ALee}, and shape observable signatures such as planetary spectra. These effects stem from the convexity of the Planck function and Jensen’s inequality \citep{1906Jensen}, highlighting the limitations of simplified homogeneous models \citep{ZhangXi2023ApJ}. While 3D models better capture these effects than 1D models, the full role of inhomogeneity in energy balance and planetary evolution requires further study.

On irradiated planets, spatial variations in incident stellar flux not only introduce energy imbalances but also drive photochemical reactions that lead to compositional inhomogeneities. 
Vertical mixing is a key process in gas giant atmospheres, disrupting their homogeneity. It is closely related to cloud and haze formation \citep{2016AA...Fromang,2018MNRAS...Ryu,2018ApJ...Zhang....1Z,2018ApJ...Zhang,2019MNRAS...Menou,WallaceOrmel2019A&A}. This mixing can generate heat, preventing cold traps \citep{Spiegel2009,Parmentier2013,2023Natur.Pelletier,2024arXivZhang}. Its strength also influences atmospheric composition, with strong mixing that depleting methane \citep{Sing2024Natur} and interacting with tidal effects \citep{Welbanks2024}. Vertical mixing, driven by convection and turbulence \citep{Zhang_2020}, affects atmospheric dynamics by influencing chemical movement and mass transport. The interaction of radiative and vertical mixing in the radiative layer is poorly understood. Gravity waves \citep{Lindzen1981, Strobel1987, Zhang_2020} and large-scale wave-induced mixing \citep{1984maph...Holton, Zhang_2020} enhance downward flux \citep{Youdin2010}, aiding upper-atmosphere chemical mixing. Although increased incident stellar flux and IR opacity contrast can enhance gas giant inhomogeneity \citep{ZhangXi2023ApJ}, vertical mixing energy transport effects are not considered.

This research integrates vertical mixing energy transport with radiative transport to better examine gas giant inhomogeneity.
Atmospheric circulation \citep{Zhang_2020} and the breaking of gravity waves \citep{Strobel1987, Zhang_2020} induce vertical mixing and downward flux \citep{Youdin2010}, forming a radiative-convective-mixing equilibrium (RCME).
Using the method from \citet{ZhangXi2023ApJ,2023ApJ...Zhang...Inhomogeneity2}, this model employs two inhomogeneous columns to compare with a homogeneous column.
We study how spatial variations in stellar flux and infrared opacity affect a planet’s atmospheric structure and cooling. These inhomogeneities—by altering radiative transfer—modify the energy balance, influence vertical mixing, and impact the overall cooling efficiency.
An increased contrast in stellar flux and IR opacity enhances inhomogeneity in the upper atmosphere. When vertical mixing is present, heat is redistributed downward, intensifying temperature variations in the lower atmosphere as well. This mixing creates a pseudo-adiabatic region, pushing the radiative-convective boundary (RCB) deeper (Papers I–III: \citealt{Zhong2024,2025ZhangApJ,2025Zhong_3}).
Although the inhomogeneity driven by the mixing can locally enhance cooling, it suppresses large-scale convection. Compared to the non-mixing case, the overall cooling capacity is reduced. Therefore, while atmospheric inhomogeneity increases under stronger flux and opacity contrasts, vertical mixing ultimately limits the planet’s ability to cool efficiently.

This study is structured as follows. 
The inhomogeneity of the gas giant can be assessed through the convexity of the Planck function alongside Jensen's Inequality, which is discussed in \S~\Ref{section_convexity}. The radiative transport equations are presented in \S~\ref{sec3_radiative_function}. Furthermore, the interplay between vertical mixing, incoming stellar flux, and IR opacity on the gas giant's inhomogeneity is evaluated in \S~\Ref{sec:Resluts}. Finally, conclusions and discussions are elaborated in \S~\ref{sec:conclusion}. The related derivations are provided in \S~Appendix~\ref{sec_appendix}.

\section{Convexity of the Planck Function and Jensen's Inequality}
\label{section_convexity}

Atmospheric inhomogeneity enhances a planet's infrared emission relative to a homogeneous atmosphere possessing the same mean brightness temperature, as demonstrated by \citet{ZhangXi2023ApJ}. For a specified frequency \(\nu\), the 
spatial and temporal average of outgoing flux is $\overline{F_\nu} = \pi \overline{B_\nu \left(T_B\right)}$, where $B_\nu$ represents the Planck function and $T_{\rm B}$ denotes the local brightness temperature, indicative of the thermal radiation of the emitting layer. 
The finite version of Jensen's inequality asserts that for a convex function $f(x)$, a collection of real numbers $X=x_1, x_2, \ldots, x_n$, and corresponding positive weights $W=w_1, w_2, \ldots, w_n$ where $\sum_{i=1}^n w_i=1$, we have the inequality:
\begin{equation}
    \sum_{i=1}^n w_{\rm i} f\left(x_i\right) \geqslant f\left(\sum_{i=1}^n w_i x_i\right) \ .
 \label{eq0:inhomo_defination_new}
\end{equation}
When each weight is $w_i={1}/{n}$, the weighted sum simplifies to the arithmetic mean. 
For any convex function, the function's expected value ($\mathbb{E}(f)$) is greater than the value of the function at the expected input.
This inequality can also be extended to a continuous version.
Jensen's inequality \citep{1906Jensen} can alternatively be expressed in another form
\begin{equation}
 \mathbb{E}[f(x)] \geq f(\mathbb{E}[x]) \ ,
 \label{eq0:inhomo_defination}
\end{equation}
guarantees $\overline{B_\nu \left(T_{\rm B}\right)} \geq B_\nu \left(\overline{T_B}\right)$, implying that non-uniform $T_{\rm B}$ distribution should emit more IR flux. This effect is further intensified for total emission, which is regulated by the Stefan-Boltzmann law ($F = \sigma T_{\rm B}^4$ with Stefan–Boltzmann constant $\sigma$).
An alternative way to state this principle is that if two planets emit the same average flux, the planet with a uniform surface should have a higher global-mean brightness temperature compared to the planet with a nonuniform surface.
In addition, the Planck function, $B_\nu$, is a convex function of temperature at every frequency. Graphically, the convexity of $B_\nu$ means that a linear combination of two brightness temperatures results in a lower outgoing flux.

Atmospheric inhomogeneity, resulting from spatial disparities in stellar flux and horizontal differences in atmospheric opacity, exerts a significant influence on cooling in RCEM framework. Heat transport in planetary atmospheres is governed by a combination of radiation, convection, large-scale circulation, and wave-driven mixing. To keep the problem tractable, we simplify the system as a set of vertical 1D columns in RCEM, with radiation and mixing dominating the upper layers and convection in the lower layers, following  \cite{ZhangXi2023ApJ}.

We examine two key sources of inhomogeneity along isobars: (1) variations in stellar flux due to changes in incident angle or orbital distance, and (2) variations in opacity driven by chemistry or horizontal transport. We first explore how internal heat flux respond nonlinearly to these factors. Then, by comparing results from inhomogeneous columns with those from an averaged, homogeneous model, we quantify the effect of atmospheric inhomogeneity on planetary cooling following  \cite{ZhangXi2023ApJ}.

\section{The Radiative Transfer}
\label{sec3_radiative_function}

In order to assess the influence of inhomogeneities on internal heat fluxes within the radiative-convective-mixing equilibrium (RCME) atmosphere, we shall utilize the second-order radiative transfer equations to derive the atmospheric structure. \S  \ref{sec2.0_radiative_transfer_functions} presents the second-order radiative transfer equations. \S \ref{sec2.1_RCE} presents the 
RCME. \S \ref{sec2.2_Method} demonstrates the methodology for solving the radiative transfer equations.

\subsection{The Second-Order Radiative Transfer Equations}
\label{sec2.0_radiative_transfer_functions}

According to \citet{Gandhi2017,2017MNRAS...Hubeny}, the second-order radiative transfer equations are:
\begin{equation}
 \frac{\partial^2 K_{\nu}}{\partial \tau_\nu^2} = \frac{\partial^2\left(f_\nu J_{\nu}\right)}{\partial \tau_\nu^2} 
 = \gamma_{\nu}^2 \left(J_{\nu}-B_{\nu}\right) \ .
\label{eq:RTE}
\end{equation}
Symbols $J_{\nu}$, $H_\nu$, and $K_{\nu}$ represent the zeroth, first, and second moments of specific intensity respectively. 
The Planck function is denoted as $B_\nu$, whereas $\gamma_{\nu}$ represents the ratio of the opacity.
Additionally, the second Eddington factor $f_{\nu}$ varies across different frequency bands ${\nu}$. The optical depth is given by $\tau_{\nu}$.

The upper boundary condition of the atmosphere is defined as
\begin{equation}
     \left.\frac{\partial\left(f_{\nu} J_{\nu}\right)}{\partial \tau_{\nu}}\right|_{\tau_\nu=0}= \gamma_{\nu} \left[g_\nu J_{\nu }-H_{\mathrm{ext}}\right]\ .
     \label{eq:upper_boundary}
\end{equation}
The second Eddington factor $g_{\nu}= H_{\nu} \left(\tau_\nu = 0\right)/ J_{\nu}\left(\tau_\nu = 0\right)$ is determined as $1/2$ for the thermal band and $-\mu$ for the visible band. 
Note that $\mu$ represents the cosine of the angle of incidence of the irradiation.
The lower boundary condition of the atmosphere is specified as 
\begin{equation}
     \left.\frac{\partial\left(f_{\nu} J_{\nu}\right)}{\partial \tau_{\nu}}\right|_{\tau_{\max }}=\gamma_{\nu} \left[\frac{1}{2}\left(B_{\nu}-J_{\nu}\right)+\frac{1}{3} \frac{\partial B_{\nu}}{\partial \tau_\nu}\right]_{\tau_{\max}} \ .
\label{eq:bottom_boundary}
\end{equation}
$\tau_{\rm max}$ is the optical depth at the bottom of the atmosphere.
The density is computed using the ideal gas law $P = {\rho k_{\mathrm{b}} T}/{\overline{m}}$, which takes into account both the average molecular mass $\overline{m}$ and the Boltzmann constant $k_{\mathrm{b}}$.

In a semi-gray atmosphere, the notation utilized in $\nu$ for frequency bands can be substituted by ``IR" to denote the thermal/IR band and ``v" for the visible band. 
In this atmosphere, the optical depth is primarily determined by the thermal opacity $\tau = \kappa_{\rm IR} \rho d z $, in relation to the density $\rho$ and altitude $z$.
In addition, the thermal band is characterized by $B_{\rm IR} = \sigma T^4/\pi$  and the visible band by $B_{\rm v} = 0$ \citep{2010A&A...Guillot}.
In the thermal band, $\gamma_{\nu}$ is described by $ \gamma_{\rm IR} = \kappa_{\rm IR}/\kappa_{\rm IR} = 1$, with $\kappa_{\rm IR}$ as the thermal opacity coefficient. According to \citet{ZhangXi2023ApJ}, $\gamma_{\nu}$ in the visible band is described by $\gamma_{\rm v} = \kappa_{\rm v}/\kappa_{\rm IR} = \alpha$ with $\kappa_{\rm v}$ as the visible opacity coefficient.
According to \citet{ZhangXi2023ApJ}, the Eddington factors are $f_{\rm IR}=K_{\rm IR}/J_{\rm IR}=1/D^{2}$ for the thermal band and $f_{\rm v}=K_{\rm v}/J_{\rm v}=1/D$ for the visible band, where the closure parameter $D$ is typically set to~2 \citep{Robinson2012Ap,ZhangXi2023ApJ}.

To simplify, we consider a direct beam with $\mu = 1$, which follows the Beer law given by $F_{\rm v}\left(\tau \right) = - F_\odot e^ {-\alpha \tau }$ \citep{ZhangXi2023ApJ} for the visible spectrum. Here, $F_\odot$ represents the normalized local incident stellar flux. Thus, at the top of the atmosphere, the stellar flux in Equation (\ref{eq:bottom_boundary}) is expressed as $H_{\mathrm{ext}} = H_{\rm v} = 4\pi F_{\odot}$. In contrast, $H_{\rm ext}$ equals zero in the thermal spectrum. For other values of $\mu$, the solution can be generalized by adjusting $F_\odot$ and $\alpha$.

\subsection{RCME}
\label{sec2.1_RCE}

The upper atmosphere must meet the RCME (Paper III),  
which should be shown as follows:
\begin{equation}
    \int_{\rm 0}^{\infty} \kappa_{\nu} \left(J_{\nu} - B_{\nu}\right)  d \nu + \frac{ g}{4 \pi} \frac{d F_{\rm conv}}{dP} + \frac{g}{4 \pi} \frac{d F_{\rm mix}}{dP} = 0 \ .
    \label{eq:energy_flux_conservation_001}
\end{equation}
Here, $F_{\rm conv}/4\pi$ and $F_{\rm mix}/4\pi$ represent the convective and mixing fluxes respectively. Both are in units of $4\sigma T_{0}^{4}$, where $T_{0}$ is the equilibrium temperature as described by \citet{ZhangXi2023ApJ}.
While Equation~(\ref{eq:energy_flux_conservation_001}) performs well at the top of the atmosphere, it becomes unstable near the bottom for large  values of $\kappa$. In deep layers, instability arises with large $\kappa$ and small $J_{\nu} - B_{\nu}$, necessitating a modified flux with $d\tau$ for stability.
Using the hydrostatic equation $d P/ dz = - \rho g dz$ with pressure $P$, altitude $z$, and density $\rho$, Equation (\ref{eq:energy_flux_conservation_001}) becomes
\begin{equation}
    \int_{\rm 0}^{\infty} \frac{1}{\gamma_{\nu}}\frac{d \left( f_{\nu} J_{\nu}\right)}{d \tau} d \nu + \frac{F_{\rm conv}}{4 \pi} +\frac{F_{\rm mix}}{4\pi} = H  \ ,
    \label{eq:energy_flux_conservation_02}
\end{equation}
where the total energy flux $ H = \sigma T_{\rm int}^4/4\pi$ (with intrinsic temperature $T_{\rm int}$) is given in units $4\sigma T_{0}^{4}$ \citep{ZhangXi2023ApJ}. In addition, a gravitational acceleration fraction represented by $g=1000 \  {\rm cm/s^2 }$. 
The notation $\kappa_{\nu}$ is expressed in terms of opacities per unit mass, specified in units of ${\rm cm^{2}/g}$ \citep{2017MNRAS...Hubeny}.
It is essential to note that within the radiative layer, $F_{\rm conv} = 0$ whereas $F_{\rm mix} \ne 0$. On the other hand, in the convective layer, $F_{\rm conv} \ne 0$ whereas $F_{\rm mix} = 0$. The equations mentioned above illustrate a state of radiative equilibrium (RE) when neither convection nor mixing is present. If convection is not present, this condition is termed radiative-mixing equilibrium (RME). Conversely, when mixing is absent, it is known as radiative-convective equilibrium (RCE).

The formulation of convective flux $F_{\rm conv}$, according to the principles of mixing length theory \citep{2013book..Kippenhahn}, is expressed by
\begin{equation}
\begin{aligned}
   F_{\rm conv} 
   \left(\nabla, T, P\right) =  
   \ F_{\rm 0} \left(\nabla-\nabla_{\rm el}\right)^{3/2} \ ,
\end{aligned}
\end{equation}
where
\begin{equation}
    F_0 = \left(\frac{g Q H_{\rm P}}{32}\right)^{1/2} \rho c_{\rm P} \left(l/H_{\rm P}\right)^2 \ , 
\end{equation}
\begin{equation}
    \nabla-\nabla_{\rm el} = \frac{A^2}{2}  + \xi - A \left(\frac{A^2}{4}+\xi\right)^{1/2}\ ,
\end{equation}
\begin{equation}
    \nabla = \frac{d\ln T}{d \ln P} = \frac{P}{T}\frac{d T}{dP}\ .
\end{equation}
Note that in the above equation, 
\begin{equation}
    A \equiv \frac{16 \sqrt{2}\sigma T^3}{\rho c_{\rm P}\left( g Q H_{\rm  P}\right)^{1/2}\left(l/H_{\rm P}\right)} \frac{\tau_{\rm el}}{1+\frac{1}{2}\tau_{\rm el}^2} \ ,
\end{equation}
and $\xi = \nabla - \nabla_{\rm ad}$.
In this scenario, $c_{\rm P}$ represents the heat capacity maintained at constant pressure, while $Q \equiv -\left(d\ln \rho/d \ln T\right)_{\rm P}$ is a parameter that equals 1 in the case of an ideal gas. The optical depth $\tau_{\rm el} = l \kappa_{\rm R}$ for a small gas segment with size $l$ is also examined, where $l$ typically approximates $\approx H_{\rm P}$, symbolizing the height of the atmospheric scale. The selection of $l$ has a minor impact on the flux detected from hot Jupiters, as convective layers are generally located far below the observable atmosphere. The elemental log-temperature gradient, indicated as $\nabla_{\rm el}$, satisfies the condition
\begin{equation}
    \nabla_{\rm el}-\nabla_{\rm ad} = A \sqrt{\nabla -\nabla_{\rm el} }\ . 
\end{equation}
By adding $\nabla$ to both sides and rearranging, $\nabla -\nabla_{\rm el}$ can be calculated from the resultant quadratic in $\sqrt{\nabla -\nabla_{\rm el}}$, thus
\begin{equation}
    \nabla-\nabla_{\rm ad} = \nabla -\nabla_{\rm el} + A\sqrt{\nabla -\nabla_{\rm el}} \ .
\end{equation}

In general, we often explore the spatial inhomogeneity across the isobars. The analysis of the normalized convective temperature structure \citep{ZhangXi2023ApJ} is articulated as follows:
\begin{equation}
T_{\mathrm{conv}}(P)=T_{\mathrm{s}}\left(\frac{P}{P_{\mathrm{s}}}\right)^{\frac{\zeta(\gamma-1)}{\gamma}} .
\end{equation}
For terrestrial planets, $T_{\mathrm{s}}$ is the normalized surface temperature at pressure $P_{\mathrm{s}}$. For giant planets, $T_{\mathrm{s}}$ is the normalized interior temperature at reference pressure $P_{\mathrm{s}}$ in the deep convective zone. The adiabatic index is $\gamma$. The symbol $\zeta$ approximates adiabatic or non-adiabatic moist processes. When $\zeta=1$, it denotes a dry adiabatic. In the solar system, moist convection can significantly reduce $\zeta$, with values ranging from $0.2$ to $0.5$ \citep{Robinson2012Ap}. 
Regarding the optical depth-pressure relationship, the convective temperature aligns with 
\begin{equation}
T_{\mathrm{conv}}^4(\tau)=T_{\mathrm{s}}^4\left(\frac{\tau}{\tau_{\mathrm{s}}}\right)^\beta \ ,
\end{equation}
assuming a constant adiabatic slope parameter $\beta=4 \zeta(\gamma-1) / n \gamma$ that governs the deep convective zone's temperature profile slope. The definition of $\beta$ differs from \citet{Robinson2012Ap} where $\beta=\zeta(\gamma-1) / \gamma$ is specified. Furthermore, the set of $\beta$ aligns with that of \citet{ZhangXi2023ApJ} and equates to $4 \beta / n$ from \citet{Robinson2012Ap}.
Thus, the adiabatic temperature gradient in this work is defined as $\nabla_{\rm ad} = d\ln T/ d \ln P = \beta /4$.

Vertical mixing flux can trigger an RME within the radiative zone. Typically, heat moves from warmer to cooler regions according to thermodynamics. Nonetheless, near the IR photosphere, vertical mixing due to atmospheric circulation or the breaking of gravity waves can redirect the flux towards the hotter zone. \citet{Youdin2010} examined the additional effects of heat flux that lead to entropy mixing, which transfers heat from the cold to the warmer regions and increases atmospheric temperatures \citep{Youdin2010,2018Leconte}. The mixing flux expression is given by
\begin{equation}\label{eq.F}
    F_{\rm mix}=-K_{\rm zz}\rho g\left(1-\frac{\nabla}{\nabla_{\rm ad}}\right) \ .
\end{equation}
The vertical mixing strength is represented by \( K_{\rm zz} \). 
Take note that the flux in the results must be normalized by $4\sigma T_{0}^4$. 

\subsection{Method}
\label{sec2.2_Method}

The radiative transfer equations transform into nonlinear forms when the RCME condition is satisfied. To tackle these non-linear equations (\ref{eq:RTE})- 
(\ref{eq:energy_flux_conservation_02}), we utilize Rybicki's method \citep{2014tsa..book...Hubeny,2017MNRAS...Hubeny,2019PhDT...Gandhi}. Rybicki's approach facilitates temperature correction within RCE/RCME models by adopting a fully linearization technique, which guarantees convergence to the selected atmospheric temperature profile. The temperature is iteratively refined until the RCE/RCME condition is met, thereby enhancing computational efficiency and stability in modeling exoplanetary atmospheres. For further information, see \S 3.3 in \citet{2017MNRAS...Hubeny}, \S 17.3 in \citet{2014tsa..book...Hubeny}, or Appendix A in Paper II. 
This work indicates that the output of the Rybicki approach matches that of the conventional complete linearization method \citep{1969ApJAuer...Mihalas}. Nevertheless, we aim to evaluate the Rybicki method's effectiveness in processing scenarios with massive frequency counts (ranging from tens to hundreds of thousands) in an upcoming study, with a specific focus on assessing chemical interactions in high-resolution atmospheric models.

\section{Results} \label{sec:Resluts}

To examine the differences between the two inhomogeneous columns and the homogeneous column, we employ the convexity properties of the Planck function in conjunction with Jensen's inequality, as elaborated in \S \ref{section_convexity}. 
This section is primarily dedicated to exploring the inhomogeneous characteristics relevant to gas giants.
These giant planets lack defined surfaces. Following \citet{ZhangXi2023ApJ}, we also assume that the atmospheric temperature is homogeneous horizontally within the deep convective layer. This assumption is predicated on the concept that convection efficiently homogenizes the horizontal temperature distribution along an isobar (a constant-pressure level) within the planet's interior. In addition, the internal flux $F_{\rm int} \ne 0$. 

The interior adiabatic is defined by $T_{\rm s}$ at a substantial reference pressure, denoted $P_{\rm s}$ (or alternatively, $\tau_{\rm s}$). This is due to the inherently arbitrary nature of selecting reference pressure and temperature on giant planets. To facilitate our discussion, we introduce a new variable: 
\begin{equation}
    K = T_{\rm s}^4 \tau_{\rm s}^{-\beta} \ , 
\end{equation} 
In this context, $K$ acts as a representative for the normalized internal temperature, or entropy, in terms of optical depth coordinates, which facilitates expressing the temperature profile in the convective layer as $T_{\rm conv}^4 \left(\tau\right) = K \tau^{\beta}$.
Furthermore, the conservation of energy flux is scaled by $4 \sigma T_{0}^4$. 
Consequently, $K$ is adjusted by dividing it by $4 \sigma T_{0}^4$.
In this work, all columns share a common deep adiabat, i.e., the representative parameter of entropy $K$ is the same.

\begin{figure*}[ht!]
    \centering
    \includegraphics[width=1.0\linewidth]{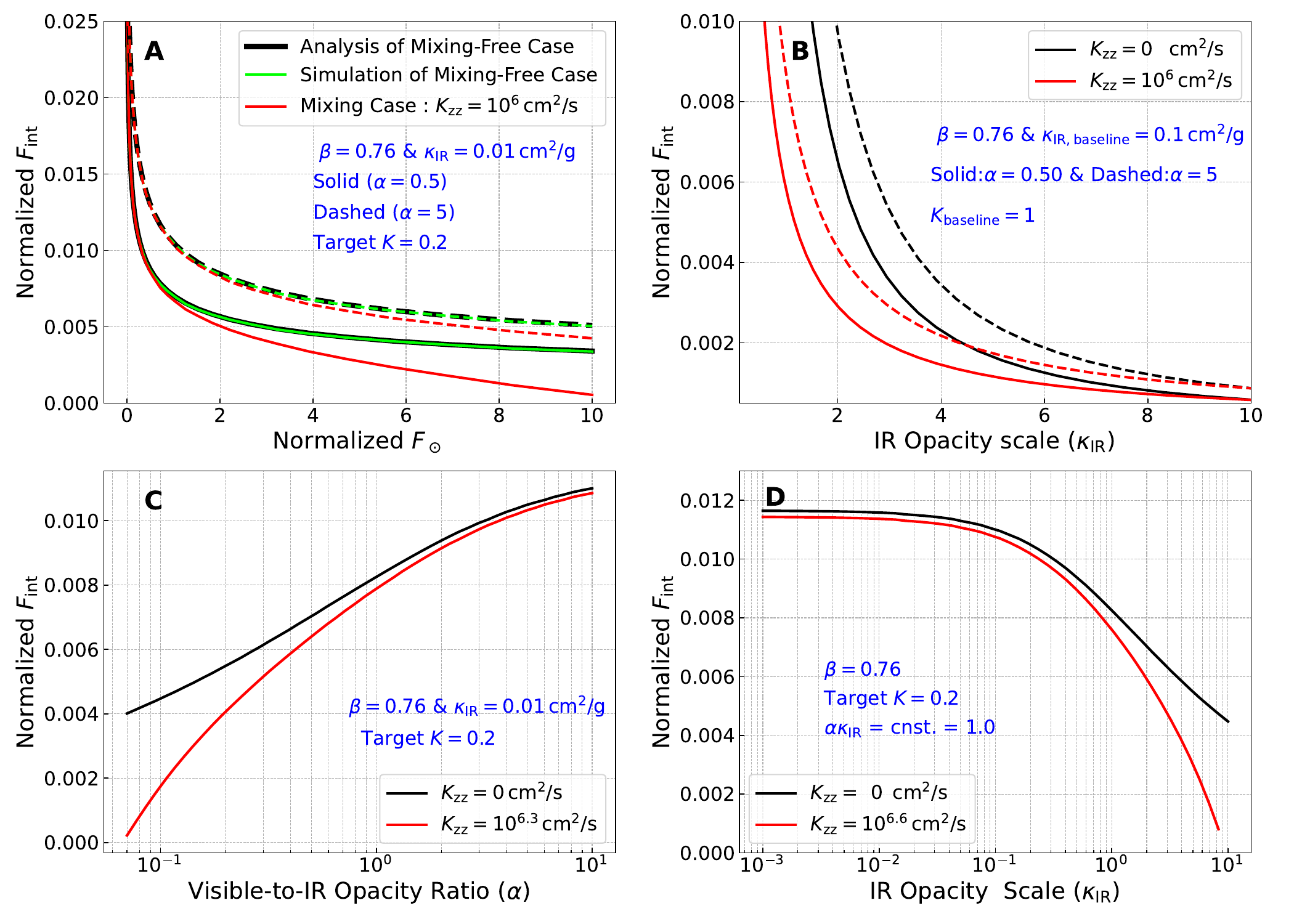}
\caption{
Panels (A)--(D) show how vertical mixing affects outgoing internal heat flux ($F_{\rm int}$) in giant planets under varying stellar flux ($F_{\odot}$), IR opacity ($\kappa_{\rm IR}$), and opacity ratio ($\alpha$).
Parameters: mixing strength $K_{\rm zz}=10^{6}-10^{6.6}\ \mathrm{cm^{2}\,s^{-1}}$, diffusivity $D=2$, adiabatic slope $\beta=0.76$, and baseline-model opacity $\kappa_{\rm IR,baseline}=0.1\ \mathrm{cm^{2}\,g^{-1}}$ and temperature parameter $K_{\rm baseline}=1$.
Black lines show no mixing; red lines show mixing. $K \propto \kappa_{\rm IR}^{-\beta}$ adjusts $K$ with opacity.
{\bf (A)} $F_{\rm int}$ vs.\ $F_{\odot}$ for $\alpha = 0.5$ (solid) and $\alpha = 5$ (dashed). The response is convex.
{\bf (B)} $F_{\rm int}$ vs.\ $\kappa_{\rm IR}$ at $F_{\odot} = 1$. The convex response varies with opacity.
{\bf (C)} $F_{\rm int}$ vs.\ $\alpha$ at $\kappa_{\rm IR} = 0.01 \mathrm{cm}^2/\mathrm{g}$, $K = 0.2$. The response is concave.
{\bf (D)} $F_{\rm int}$ vs.\ $\kappa_{\rm IR}$ with fixed visible opacity ($\alpha$ constant). The convex response mirrors Panel (B).
}
\label{fig1:}
\end{figure*}

\subsection{Response Functions of Outgoing Internal Heat Flux in Mixing Case}
\label{Sec4.1}

Initially, in the context of non-mixing and mixing scenarios, the RE / RME column model will be utilized in conjunction with condition $\nabla_{\rm ad} = \beta/4$ and the temperature at the radiative-convective boundary (RCB) $T_{\rm rad} \left(\tau_{\rm rcb}\right) = T_{\rm conv} \left(\tau_{\rm rcb}\right)$ to  ascertain $F_{\rm int}\left(F_{\odot}, \kappa_{\rm IR}, \alpha, K_{\rm zz} \right)$. 
For the normalized interior entropy proxy denoted as \( K \equiv T_s^4 \tau_s^{-\beta} \), we have selected a moderate value of \( K = 0.2 \). This particular value corresponds to physically plausible conditions found in the deep atmospheres of giant planets.
Furthermore, we investigate the impact of atmospheric inhomogeneity concerning ($F_{\odot}, \kappa_{\rm IR}, \alpha, K_{\rm zz}$ ) on the heat flux emitted at the upper boundary of the atmosphere. 
The findings are presented in Figure \ref{fig1:}. 
We investigated the impact of vertical mixing on the outgoing radiative flux through four scenarios: varying the incident stellar flux $F_{\odot}$ while keeping opacity and mixing strength constant (Figure~\ref{fig1:}~A); varying the IR opacity $\kappa_{\rm IR}$ with fixed stellar flux and visible-to-IR opacity ratio (Figure~\ref{fig1:}~B); varying the visible opacity by adjusting the ratio $\alpha$ with stellar flux and IR opacity held constant (Figure~\ref{fig1:}~C); and varying the IR opacity $\kappa_{\rm IR}$ while maintaining constant stellar flux and visible opacity (Figure~\ref{fig1:}~D). 
We focus on inhomogeneities along isobars. For simplicity, we assume a direct stellar beam with $\mu = 1$, and model the flux using Beer's law: $F_{\rm v}(\tau) = -F_{\odot} e^{-\alpha \tau}$, where $F_{\odot}$ is the normalized local incident flux. Solutions for other $\mu$ values can be obtained by rescaling $F_{\odot}$ and $\alpha$ \citep{ZhangXi2023ApJ}.

To assess the accuracy of our model, we initially compared our mixing-free numerical simulations with the analytical solutions outlined in \citet{ZhangXi2023ApJ}. Panel~(A) in Figure~\ref{fig1:} illustrates the link between the intrinsic energy output of the planet ($F_{\rm int}$) and the incoming stellar flux ($F_\odot$), where the black lines indicate the analytical solutions of \citet{ZhangXi2023ApJ}, and the lime lines depict our numerical results (dashed line for $\alpha = 5$, solid line for $\alpha = 0.5$). 
The case of $\alpha = 0.5$ represents atmospheres with weak visible-wavelength absorption, allowing stellar radiation to penetrate deeply. In contrast, the case of $\alpha = 5$ captures strong upper-atmosphere absorption from species like TiO or hazes, which generates temperature inversions. These two limits bracket the observed diversity from non-inverted to strongly inverted exoplanet atmospheres, ensuring our proof-of-concept spans the key physical regimes.
This figure shows that our numerical solutions align closely with the analytical results, affirming the credibility of our numerical method. Although a similar level of agreement was observed in simulations without mixing shown in panels B, C, and D of Figure~\ref{fig1:}, analytical solutions were not included in these panels to maintain clarity.

Vertical mixing affects the $F_{\mathrm{int}}$–$F_{\odot}$ relationship, shown by the red curves in Figure~\ref{fig1:}(A). Analyzing this requires acknowledging temperature changes influenced by RME or energy conservation principles. In the semi-gray atmosphere, Equation (\ref{eq:energy_flux_conservation_001}) becomes 
\begin{equation}
    \kappa_{\rm IR} J_{\rm IR} - \kappa_{\rm IR} B + \kappa_{\rm v} J_{\rm v} + \frac{ g}{4\pi} \frac{d F_{\rm mix}}{dP} = 0,
    \label{eq:sect4.2_kappa_th}
\end{equation}
where $B = \sigma T^4 / 4\pi$, and the total flux satisfies:
\begin{equation}
    H = H_{\rm IR} + H_{\rm v} + \frac{F_{\rm mix}}{4\pi} = \frac{\sigma T_{\rm int}^4}{4\pi}.
    \label{eq:energy_flux_conservation_01}
\end{equation}

According to the findings presented in \citet{2010A&A...Guillot}, the zero moment intensity of the visible band is characterized as follows:
\begin{equation}
    J_{\rm v} = J_{\rm v}(0) e^{-\alpha \tau} = \frac{H_{\rm v}(0)}{\mu} e^{-\alpha \tau},
\end{equation}
with  optical depth relates to pressure via $d\tau_{\nu} = \kappa_{\nu} dP / g$, yielding:
\begin{equation}
    \frac{d H_{\nu}}{dP} = \gamma_{\nu} H_{\rm v}.
\end{equation}
The thermal intensity is:
\begin{equation}
    J_{\rm IR} = J_{\rm IR}(0) + \frac{1}{f_{\rm IR}} \int_{0}^{P} \frac{\kappa_{\rm IR}}{ g} H_{\rm IR} dP',
    \label{eq:sec4.2_Jth}
\end{equation}
where $f_{\rm IR} = K_{\rm IR} / J_{\rm IR}$. Substituting $H_{\rm IR} = \sigma T_{\rm int}^4 / 4\pi - H_{\rm v} - F_{\rm mix} / 4\pi$ into Equation (\ref{eq:sec4.2_Jth}) gives:
\begin{equation}
    J_{\rm IR} = J_{\rm IR}(0) + \frac{1}{f_{\rm IR}} \int_{0}^{P} \frac{\kappa_{\rm IR}}{ g} \left( \frac{\sigma T_{\rm int}^4}{4\pi} - H_{\rm v} - \frac{F_{\rm mix}}{4\pi} \right) dP'.
    \label{eq:sec4.2_Jth_updated}
\end{equation}

Combining these with Equation (\ref{eq:sect4.2_kappa_th}), the thermal emission becomes:
\begin{equation}
    B = \frac{\sigma T^4}{4\pi} = J_{\rm IR} + \frac{\alpha}{\mu} H_{\rm v}(0) e^{-\alpha \tau} + \frac{g}{4\pi \kappa_{\rm IR}} \frac{d F_{\rm mix}}{dP}.
    \label{eq:sec4.2_B}
\end{equation}
At the upper boundary, $H_{\rm v}(0) = \sigma T_{\rm irr}^4 / 4\pi$ and $H_{\rm IR}(0) = H - H_{\rm v}(0) - F_{\rm mix}(0) / 4\pi$. Using $f_{\rm IR} = 1/D^2$ and $g_{\rm IR} = 1/D$, the normalized radiative temperature is:
\begin{equation}
    T_{\rm rad}^4(\tau) =  S\left(\tau \right)  + F_{\rm int} \left( D + D^2 \tau \right) + T_{\rm mix}^4 \ ,
\label{eq:sec4.2_Trad}
\end{equation}
where 
\begin{equation} \label{eq:S}
\begin{aligned}
 S\left(\tau \right) = & \frac{\partial F_{\mathrm{v}}}{\partial \tau}-D^2 \int_0^\tau F_{\mathrm{v}} d \tau-D F_{\mathrm{v}}(0)\\
 = & F_{\odot} \left[ D +\frac{D^2 \mu}{\alpha} + \left( \frac{\alpha}{\mu} - \frac{D^2 \mu}{\alpha} \right) e^{-\alpha \tau} \right]   \ .   
\end{aligned} 
\end{equation}

The mixing term is:
\begin{equation}
    T_{\rm mix}^4 = \frac{1}{T_0^4} \left[ \frac{g}{\kappa_{\rm IR}} \frac{d F_{\rm mix}}{dP} - \frac{D^2 \kappa_{\rm IR}}{ g} \int_{0}^{P} F_{\rm mix} dP' \right].
\end{equation}
The term $-F_{\rm mix}(0) / g_{\rm IR}$ is negligible due to the low density at the upper boundary. Equation (\ref{eq:sec4.2_Trad}) aligns with Equations (14) and (16) in \citet{ZhangXi2023ApJ} when $T_{\rm mix}^4 = 0$, confirming the consistency without mixing.

We can use the two-stream method to analyze the normalized $F_{\rm rad}^{+}$. 
As shown in Appendix~\ref{sec_appendix}, the upward radiative flux is
\begin{equation}
\begin{aligned}
    F_{\rm rad}^{+} (\tau)  = & \frac{1}{2}\left[\frac{S(\tau)}{D}-\frac{1}{D} \frac{\partial F_{\mathrm{v}}}{\partial \tau}-F_{\mathrm{v}}(\tau)+\left(2+D \tau \right) F_{\mathrm{int}}\right] \\
    & -\frac{1}{2} \int_{0}^{\tau} T_{\rm mix}^4 e^{-D(\tau-t)}dt \ .
\end{aligned}
\label{eq:appendix_Frad}
\end{equation}
In addition, the  normalized $F_{\rm conv}^{+} (\tau)$  should satisfy 
\begin{equation}
\begin{aligned}
F_{\mathrm{conv}}^{+}(\tau) & =\frac{1}{2} \int_\tau^{\infty} T_{\mathrm{s}}^4\left(\frac{t}{\tau_{\mathrm{s}}}\right)^\beta e^{-D(t-\tau)} d t \\
& =\frac{T_{\mathrm{s}}^4 \tau_{\mathrm{s}}^{-\beta} e^{D \tau}}{2 D^{\beta+1}} \Gamma(\beta+1, D \tau) \ .
\end{aligned}
\end{equation}

To moderately assess the response function of $F_{\rm int}$ related to the atmospheric composition of gas giants with vertical mixing, the governing criterion at RCB should be observed \citep{ZhangXi2023ApJ}:
\begin{equation}
    T_{\rm rad}^{4} (\tau_{\rm rcb}) = T_{\rm conv}^4(\tau_{\rm rcb}) \ ,
\end{equation}
\begin{equation}
    F_{\rm rad}^{+} (\tau_{\rm rcb}) = F_{\rm conv}^{+}(\tau_{\rm rcb}) \ .
\end{equation}

We can get the expressions for $K$ and $F_{\rm int}$:
\begin{equation}
  K = \frac{ \frac{S(\tau_{\rm rcb})}{1+D\tau_{\rm rcb}}
+\left.\frac{\partial F_{\mathrm{v}}}{\partial \tau} \right|_{\tau_{\rm rcb}}+DF_{\mathrm{v}}(\tau_{\rm rcb})+ j_{\rm mix}}{\tau_{\rm rcb}^{\beta} \left(1+\frac{1}{1+D\tau_{\rm rcb}}\right) - D^{-\beta}e^{D\tau_{\rm rcb}}\Gamma \left(\beta+1, D \tau_{\rm rcb}\right)} \ , 
 \label{eq:sec4.2_final_K}
\end{equation}
and 
\begin{equation}
    \begin{aligned}
        F_{\rm int}= & \,\,\, \frac{K \tau_{\rm rcb}^{\beta}-S(\tau_{\rm rcb}) + T_{\rm mix}^4}{D + D^2 \tau_{\rm rcb}}\\
                   = & \,\,\, \frac{\left(1-j_{\rm r}\right)S\left(\tau_{\text {rcb }}\right)}{\left(D+D^2 \tau_{\text {rcb }}\right)j_{\rm r}}+\left.\frac{1}{D j_{\rm r}} \frac{\partial F_{\mathrm{v}}}{\partial \tau}\right|_{\tau_{\text {rcb}}}+ \frac{F_{\mathrm{v}}\left(\tau_{\text {rcb }}\right) }{j_{\rm r}} \\
                    & + \left[\frac{j_{\rm mix}}{Dj_{\rm r}} -\frac{T_{\rm mix}^4 (\tau_{\rm rcb})}{D + D^2\tau_{\rm rcb}} \right] \ ,
    \end{aligned}
    \label{eq:sec4.2_final_Fint}
\end{equation}
where 
\begin{equation}
    j_{\rm mix} = \frac{2+D\tau_{\rm rcb}}{1+D\tau_{\rm rcb}}  T_{\rm mix}^4 (\tau_{\rm rcb}) +  D \int_{0}^{\tau_{\rm rcb}} T_{\rm mix}^4 e^{-D(\tau-t)}dt \ , 
\end{equation}
\begin{equation}
    j_{\rm r} = 2+D \tau_{\rm rcb}-\frac{\left(1+D \tau_{\text {rcb }}\right) e^{D \tau_{\text {rcb }}}}{(\tau_{\rm rcb}D)^{\beta}} \Gamma\left(\beta+1, D \tau_{\text {rcb }}\right) \ .
\end{equation}
In this context, \( j_{\text{mix}} \) represents the incremental energy flux introduced exclusively by vertical mixing, whereas \( j_{\text{r}} \) serves as a dimensionless weighting factor derived purely from the radiation field under mixing-free conditions.
In the absence of mixing (i.e., \( j_{\rm mix} = 0 \), \( T_{\rm mix} = 0 \)), Equations (\ref{eq:sec4.2_final_K}) through (\ref{eq:sec4.2_final_Fint}) correspond with the findings of \citet{ZhangXi2023ApJ}.
The relationship between $F_{\odot}$ and $F_{\rm int}$ is illustrated in Panel~(A) of Figure~\ref{fig1:}.
The black and lime curves indicate that, without vertical mixing, the normalized $F_{\rm int}$ decreases with increasing incident stellar flux before gradually reaching a plateau. This plateau marks a critical point beyond which further increases in stellar flux have minimal impact on the position of the RCB. This is because the stellar flux reaching the RCB becomes nearly zero, according to Equation~(\ref{eq:S}). Nevertheless, the introduction of vertical mixing modifies this previously stable pattern.


Vertical mixing acts like a mechanical greenhouse, warming the lower atmosphere and pushing the radiative-convective boundary (RCB) deeper. This disrupts the previously stable structure and reduces the internal heat flux $F_{\rm int}$. The convex shape of the $F_{\rm int}$–$F_\odot$ curve—especially at $\alpha = 0.5$—demonstrates an accelerated decline in $F_{\rm int}$ due to mixing, consistent with Equation~(\ref{eq:sec4.2_final_Fint}) assuming constant diffusivity $K$. 
This convexity  reflects the nonlinear response of internal flux to stellar forcing. As shown in Figure~\ref{fig1:}(A), this behavior persists with or without vertical mixing.
According to Equation~(\ref{eq:sec4.2_Trad}), the positive contribution of $T_{\rm mix}^4$ raises the radiative temperature $T_{\rm rad}^4$, thereby increasing energy input to the deep atmosphere and enhancing the suppression of $F_{\rm int}$. As $F_\odot$ increases, mixing becomes more efficient and the RCB shifts inward, further amplifying this effect.

The strength of vertical mixing is closely tied to the optical ratio $\alpha = \kappa_{\rm v} / \kappa_{\rm IR}$, with $\kappa_{\rm IR}$ fixed. When $\alpha$ is small, the visible opacity $\kappa_{\rm v}$ is low, allowing stellar radiation and mixing-driven flux to penetrate more deeply. This strengthens vertical mixing and drives the RCB further downward, often making $F_{\rm int}$ negligible. Equation~(\ref{eq:sec4.2_final_Fint}) also shows that the negative term involving $T_{\rm mix}^4(\tau_{\rm rcb})$ becomes dominant in the expression ${j_{\rm mix}}/{D j_{\rm r}} - {T_{\rm mix}^4 (\tau_{\rm rcb})}/({D + D^2\tau_{\rm rcb}})$ as $\tau_{\rm rcb}$ increases, further reducing $F_{\rm int}$.

Figure~\ref{fig1:}(A) and (C) illustrate the role of $\alpha$. In panel (A), the solid red line shows a steeper decline in $F_{\rm int}$ than the dashed red line, highlighting the role of vertical mixing in different $\alpha$. Panel (C) shows that at lower $\alpha$, mixing more strongly suppresses $F_{\rm int}$, while at higher $\alpha$, the effect weakens. All curves exhibit concave shapes, indicating increased overall cooling efficiency with lower $\alpha$. 
This also alters the curvature of the response: while \citet{ZhangXi2023ApJ} found $F_{\rm int}$ to be concave in $\alpha$ without mixing, vertical mixing increases this concavity
(\(\partial^2 F_{\rm int}/\partial \alpha^2<0\)), as seen in Figure~\ref{fig1:}(C).
Thus, decreasing visible opacity enhances the thermal efficiency of vertical mixing in reducing internal heat loss.

To examine how infrared opacity (\(\kappa_{\rm IR}\)) influences internal heat flux (\(F_{\rm int}\)), we consider a planet receiving uniform stellar flux (\(F_\odot = 1\)). Along an isobar, the internal temperature parameter follows \(K \propto \kappa_{\rm IR}^{-\beta}\), based on the relation \(K \kappa_{\rm IR}^\beta = \text{const}\). We use a reference state (baseline model) with \(\kappa_{\rm IR,baseline} = 0.1\, \mathrm{cm}^2/\mathrm{g}\) and \(K_{\rm baseline} = 1\).
Figure~\ref{fig1:}(B) presents the results. In the absence of mixing (black lines), \(F_{\rm int}\) decreases rapidly as \(\kappa_{\rm IR}\) increases. This trend reflects the reduction in \(K\), which lowers entropy and pushes the radiative-convective boundary (RCB) deeper into the atmosphere \citep{ZhangXi2023ApJ}.

When vertical mixing is included (red lines), both \(F_{\rm int}\) and \(K\) decrease further, and the RCB shifts to even deeper layers. This indicates enhanced energy transport by mixing. However, as \(\kappa_{\rm IR}\) increases, the atmosphere becomes denser, and the effectiveness of mixing is reduced. According to Equation~(\ref{eq:sec4.2_B}), \(\kappa_{\rm IR}\) appears in the denominator of the mixing term, which means that higher opacity suppresses the energy contribution from vertical mixing. As a result, in high-opacity atmospheres, mixing becomes less efficient at regulating \(F_{\rm int}\), further limiting the planet’s ability to cool.
This result highlights the dual role of \(\kappa_{\rm IR}\): it controls both radiative cooling and the efficiency of mechanical energy redistribution through vertical mixing.
Additionally, as shown in Figure~\ref{fig1:}(B), the internal flux exhibits a convex response
to \(\kappa_{\rm IR}\), reinforcing the nonlinear sensitivity of $F_{\rm int}$ to opacity structure in the atmosphere.

With visible opacity fixed at $\alpha \kappa_{\rm IR} = 1.0$ and only IR opacity varied, the results are presented in Figure~\ref{fig1:}~(D). 
The rise in IR opacity greatly diminishes the cooling of planetary interiors as $F_{\rm int}$ decreases, further intensified by vertical mixing.
As illustrated, the difference between the red solid line (including vertical mixing) and the black solid line (excluding vertical mixing) grows with increasing IR opacity scale. Vertical mixing (red line) strengthens the convexity of the relationship between $F_{\rm int}$ and the IR opacity scale. 
Therefore, when vertical mixing is included, the combination of inhomogeneous stellar flux and atmospheric opacity leads to reduced interior cooling, as indicated by a significant decrease in $F_{\rm int}$ compared to the non-mixing case.

\subsection{Impact of Inhomogeneities on outgoing internal flux in Mixed Environments} \label{sec:floats}

\begin{figure*}
    \centering
    \includegraphics[width=1.0\linewidth]{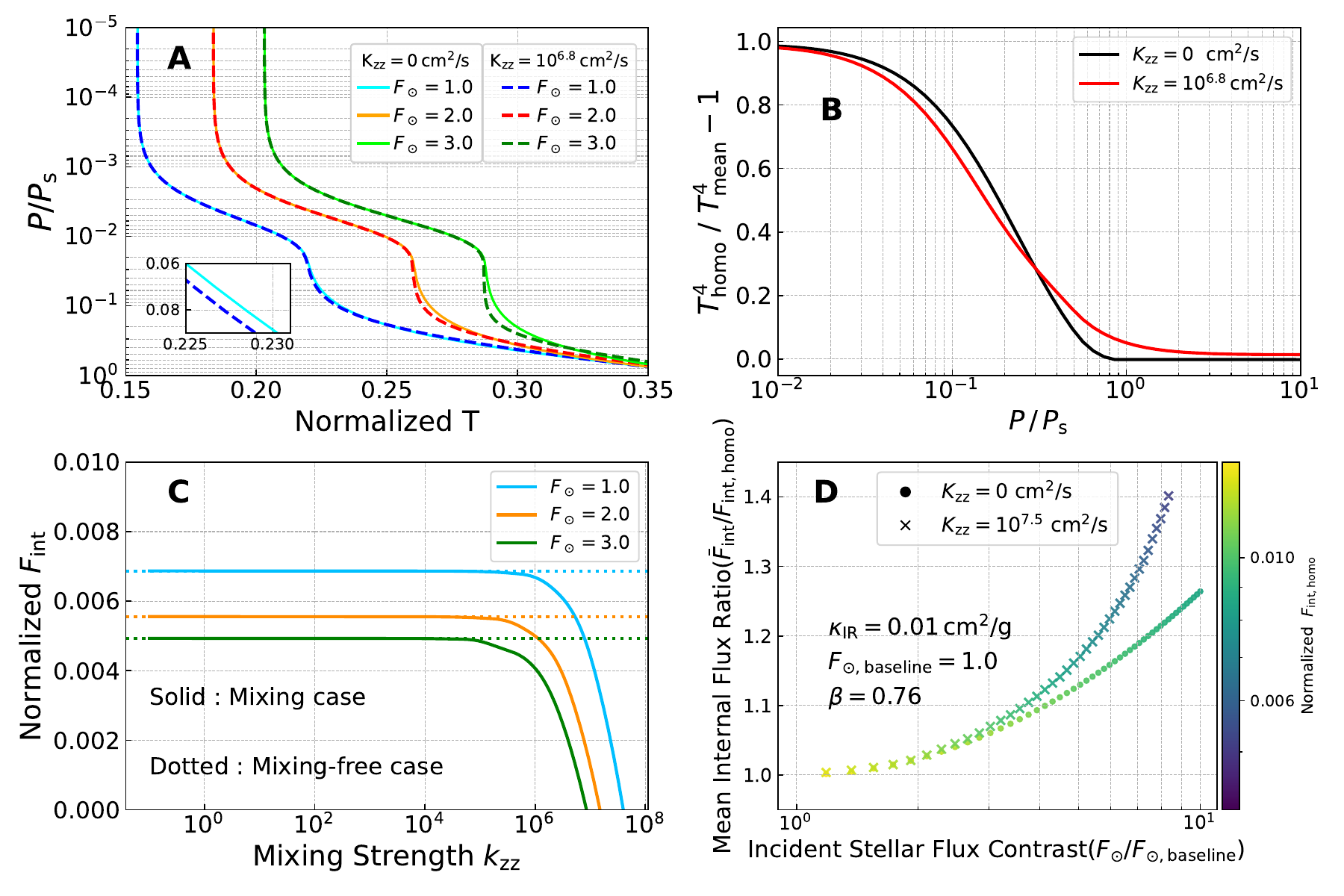}
    \caption{
Panels (A)--(D) show vertical mixing effects on a giant planet's atmosphere with $\alpha = 0.5$, comparing two inhomogeneous RCME models and one homogeneous model. Parameters: $K = 0.2$, $D = 2$, $\kappa_{\rm IR} = 0.1 \mathrm{cm}^2/\mathrm{g}$. 
Dashed/solid pairs (blue--cyan, red--orange, green--lime) denote $F_{\odot}=1.0,2.0$ and $3.0$ for mixing (no-mixing).
Dashed or red lines show mixing ($K_{\rm zz} = 10^{6.8} \mathrm{cm}^2/\mathrm{s}$ in (A)--(B), $10^{7.5} \mathrm{cm}^2/\mathrm{s}$ in (D)).
{\bf (A) } Temperature vs.\ pressure ratio ($P/P_{\rm s}$) for varying $F_{\odot}$. An inset highlights mixing onset.
{\bf (B)} Temperature ratio ($T_{\rm homo}^4 / T_{\rm mean}^4 - 1$) vs.\ pressure ratio ($P/P_{\rm s}$).
{\bf (C)} Normalized internal flux ($F_{\rm int}$) vs.\ mixing strength ($K_{\rm zz}$). Dotted lines show no mixing.
{\bf (D)} Internal flux ratio ($\bar{F}_{\rm int} / F_{\rm int,homo}$) vs.\ $F_{\odot}$ with fixed opacity. 
Dots show no mixing; crosses show mixing. The colorbar indicates normalized $F_{\rm int,homo}$.}
    \label{fig2:}
\end{figure*}

The Planck function and $T_{\rm B}^4$ are convex, so Jensen’s inequality shows $\langle B_\nu(T_{\rm B})\rangle \ge B_\nu(\langle T_{\rm B}\rangle)$ and $\langle F_\nu (T_{\rm B})\rangle \ge F_\nu(\langle T_{\rm B}\rangle)$ at each frequency $\nu$. Due to temperature and opacity changes with longitude, latitude, and the day-night cycle, $T_{\rm B}$ is uneven across infrared windows and absorption bands. Typically, we can divide the planet into $N$ radiative–convective columns, calculate each column’s spectral flux from local insolation and $\kappa_\nu$, and average to get $\langle F_\nu\rangle$. Comparing this with a uniform column gives the per-frequency radiative excess $\Delta F_\nu$ and its integral $\Delta F$, whose leading magnitude relates to $\mathrm{Var}(T)$.

This section explores the effects of vertical mixing on atmospheric inhomogeneity in gas giants using a simplified two-column model. Each column receives a different amount of incident stellar flux, emits a distinct internal heat flux, and has a unique atmospheric opacity. By comparing the average internal flux from this inhomogeneous setup with that from a homogeneous case (with uniform stellar flux and opacity), we assess how radiative differences shape planetary cooling.

The outgoing atmospheric flux governs a planet’s cooling and overall energy equilibrium, but it is not uniform. Uneven stellar irradiation (\S~\ref{sec_4.2.1_different Fodot}) produces strong spatial contrasts, under which vertical mixing plays a crucial role in redistributing energy across atmospheric layers. Differences in internal heat flux (\S~\ref{sec_4.2.2_different Fint}) influence how efficiently energy escapes from the planetary interior, and under these conditions vertical mixing modifies the inhomogeneous temperature structure at different levels. Spatial variations in opacity (\S~\ref{sec_4.2.3_different kappa}), driven by clouds, hazes, chemical composition, and circulation, control the efficiency of radiative transfer. Together, these processes give rise to inhomogeneous outgoing fluxes that alter both local temperature profiles and the global transport of energy.
Our method compares two inhomogeneous atmospheric columns with a corresponding homogeneous one, where an increase in “contrast” denotes stronger departures from uniformity in irradiation, opacity, or vertical mixing.

\subsubsection{The effect of the inhomogeneous incident stellar flux }\label{sec_4.2.1_different Fodot}

As discussed in \S~\ref{Sec4.1}, the inhomogeneity of incident stellar flux, caused by variations in incident angle or orbital distance, can be studied by scaling the incident flux and the parameter \(\alpha\).
Figure~\ref{fig2:}(A) illustrates how vertical mixing affects temperature distribution in the atmosphere for different stellar flux values. Temperature profiles for inhomogeneous columns are shown for \(F_{\odot} = 1.0\) ( blue-cyan columns) and \(F_{\odot} = 3.0\) (green-lime columns), while the homogeneous case is represented by \(F_{\odot} = 2.0\) (red-orange column). 
We adopt standard parameters: $\beta\approx 0.76$ for a dry-adiabatic H$_2$ atmosphere, $D=2$ under the hemi-isotropic approximation, $\mu=1.0$, and $\alpha=0.5$.
A lower \(\alpha\) allows more external energy to penetrate through visible wavelengths, leading to a steady increase in temperature with pressure \citep{2010A&A...Guillot}. The mechanical greenhouse effect induced by vertical mixing partially suppresses convection.

The effect of mixing is shown in Figure~\ref{fig2:}(A), where solid lines represent cases without mixing, and dashed lines show the results with mixing. For the same \(F_{\odot}\), temperature changes are determined by the relationship between internal flux (\(F_{\rm int}\)) and mixing intensity, as illustrated in Figure~\ref{fig2:}(C). Without mixing, the \(F_{\odot} = 1.0\) case (blue dotted line) results in a higher internal flux, while the homogeneous case (\(F_{\odot} = 2.0\), orange column) aligns more closely with the \(F_{\odot} = 3.0\) case (green column). 
As mixing intensity increases, $ F_{\rm int} $ decreases rapidly in a nonlinear manner.
At steady mixing, the homogeneous column’s \( F_{\rm int} \) approaches that of the higher flux inhomogeneous column. However, intense mixing amplifies \( F_{\rm int} \) inhomogeneity, with \( F_{\rm int} \) approaching negligible values, except when \( F_{\rm int} = 0 \).

Equations~(\ref{eq:energy_flux_conservation_001})--(\ref{eq:energy_flux_conservation_02}) show that in the RCME, vertical mixing offsets the temperature drop caused by a lower internal heat flux, \( F_{\rm int} \), stabilizing the temperature. However, vertical mixing often heats the system more than the \( F_{\rm int} \) drop cools it, potentially raising the temperature.
Vertical mixing also slows the cooling of the planet’s interior by speeding up heat loss as \( F_{\rm int} \) decreases. Figure~\ref{fig2:}(C) shows that a higher stellar flux, \( F_\odot \), means less mixing is needed to reduce \( F_{\rm int} \). So, a high \( F_\odot \) cuts \( F_{\rm int} \) more at the same mixing level, making temperature rises more likely.
Thus, the inhomogeneous column at \( F_\odot = 3.0 \) in Figure~\ref{fig2:}(A) has stronger bottom heating due to larger \( F_{\rm int} \) changes compared to the homogeneous column (orange dashed line). At lower \( F_\odot \) (blue and orange dashed lines), temperature changes are smaller, and bottom temperatures are more uniform.
Still, the RCB shifts to a deeper level.

We apply the Jensen’s inequality in Equation~(\ref{eq0:inhomo_defination}) of \S~\ref{section_convexity} to evaluate inhomogeneity,  which is achieved by comparing  \(T_{\rm mean}^4\) (actual mean temperature of an inhomogeneous planet) with \(T_{\rm homo}^4\) (uniform temperature of a hypothetical homogeneous planet with globally averaged flux and opacity).
Figure~\ref{fig2:}(B) shows the temperature ratio (\( T_{\rm homo}^4 / T_{\rm mean}^4 \), always above 1) drops as pressure increases. This means the outer atmosphere has more temperature inhomogeneity, while the interior is more uniform.
Vertical mixing changes this trend. In the upper atmosphere, the temperature ratio with mixing is almost the same as without mixing. But in higher-pressure regions, mixing first lowers temperature variability up to a pressure of about 0.5. Beyond this point, variability rises compared to cases without mixing.

We studied how the stellar flux contrast ($F_{\odot}/F_{\odot,{\rm baseline}}$) affects the mean internal flux ratio (\( \bar{F}_{\rm int}/F_{\rm int,homo} \)), as shown in Figure~\ref{fig2:}~(D). The first set of inhomogeneous columns has a constant baseline stellar flux ($F_{\odot,{\rm baseline}}$). The second set varies due to changing stellar flux.
A higher $F_{\odot}/F_{\odot,{\rm baseline}}$ indicates more uneven \( F_{\odot} \). Figure~\ref{fig1:}(A) shows \( F_{\rm int} \) is a convex function of \( F_\odot \). Using the Jensen’s inequality in Equation~(\ref{eq0:inhomo_defination}), we find that for two-column cases, greater \( F_{\odot} \) unevenness increases the convex function’s mean value. Thus, in Figure~\ref{fig2:}~(D), the ratio \( \bar{F}_{\rm int}/F_{\rm int,homo} \) rises with higher $F_{\odot}/F_{\odot,{\rm baseline}}$.

Vertical mixing strengthens this trend. It increases the convexity of \( F_{\rm int} \), raising the ratio \( \bar{F}_{\rm int}/F_{\rm int,homo} \) (crosses in Figure~\ref{fig2:}~(D)) above the no-mixing case (dots). This gap widens as $F_{\odot}/F_{\odot,{\rm baseline}}$ grows. In a homogeneous column with average stellar flux, mixing boosts internal flux emission by over 1.4 times compared to the uniform case.

However, vertical mixing slows global planetary cooling compared to no mixing. Without mixing, uneven stellar flux boosts cooling by increasing emission \citep{ZhangXi2023ApJ}. Uneven flux also makes mixing more efficient, enhancing local cooling. Yet, total internal flux emission stays lower with mixing.

Mixing efficiency depends on the homogeneous column’s internal flux (\( F_{\rm int,homo} \)). In Figure~\ref{fig2:}(D), dot and cross shading shows \( F_{\rm int,homo} \) values, which peak near high stellar flux regions in Figure~\ref{fig2:}(C). More uneven stellar flux leads to greater \( F_{\rm int} \) reduction in high-flux areas, as shown by darker shading in Figure~\ref{fig2:}(D).
While \( \bar{F}_{\rm int}/F_{\rm int,homo} \) increases with flux contrast, vertical mixing reduces the mean internal flux (\( \bar{F}_{\rm int} \)), slowing global cooling compared to no mixing.

\begin{figure}
    \centering
    \includegraphics[width=1.0\linewidth]{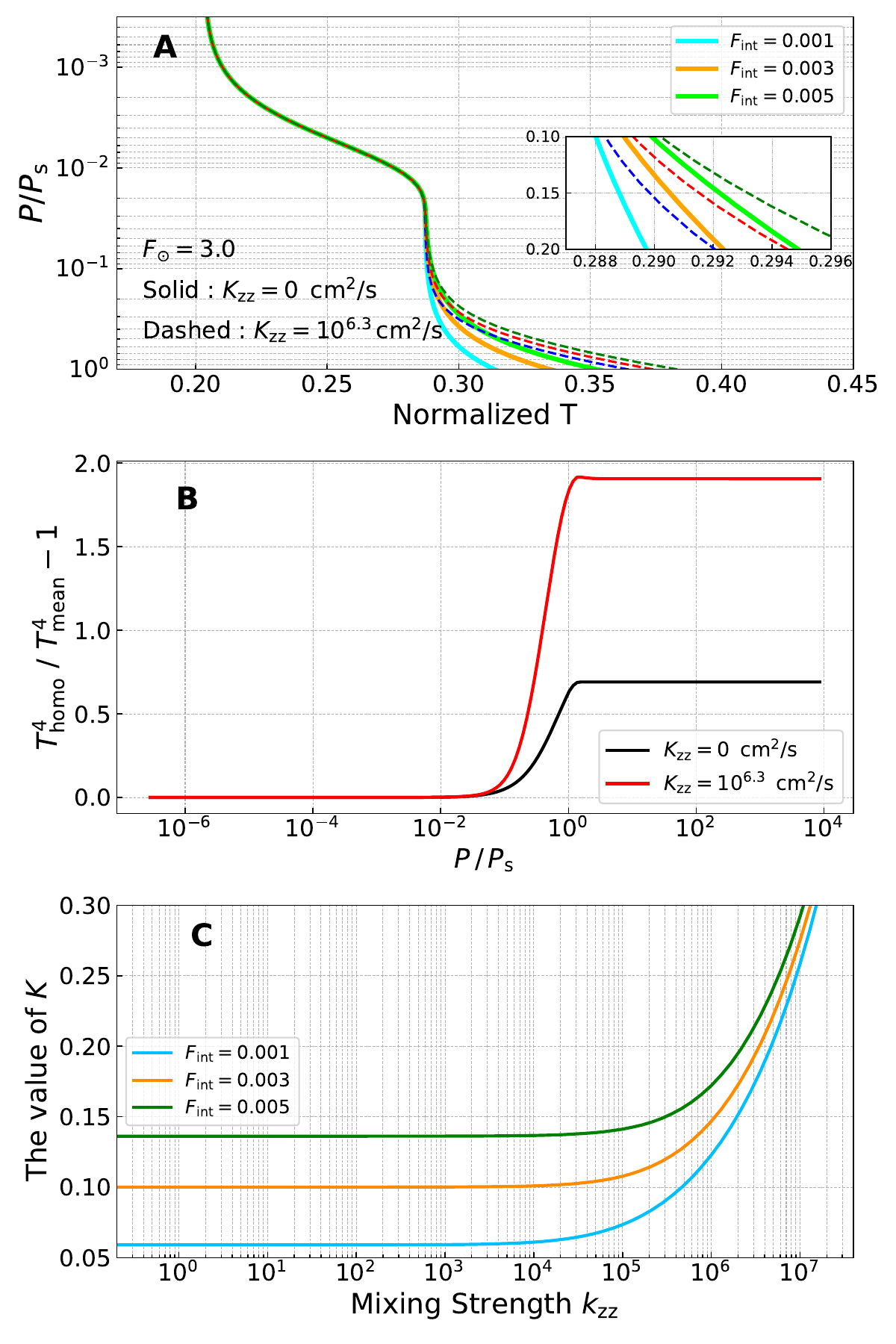}
    \caption{
    Panels (A)–(C) show temperature and mixing effects for two inhomogeneous and one homogeneous column (\( \alpha = 0.5 \), \( K = 0.2 \), \( D = 2 \), \( \kappa_{\rm IR} = 0.1 \, \text{cm}^2/\text{g} \), \( F_\odot = 1.0 \)). Blue, green, and orange lines represent \( F_{\rm int} = 0.001 \), 0.005, and 0.003, respectively. Solid lines indicate no vertical mixing; dashed or red lines show mixing (\( K_{\rm zz} = 10^{6.3} \, \text{cm}^2/\text{s} \) in (A)).  
\textbf{(A)} Temperature vs. pressure profiles, with an inset zooming in on details.  
\textbf{(B)} Temperature ratio (\( T_{\rm homo}^4 / T_{\rm mean}^4 - 1 \)) vs. pressure ratio (\( P/P_{\rm s} \)). Black lines: no mixing; red lines: mixing.  
\textbf{(C)} Internal temperature parameter (\( K \)) vs. vertical mixing strength (\( K_{\rm zz} \)).  
    }
    \label{fig3}
\end{figure}

\subsubsection{The effect of the inhomogeneous outgoing fluxes }\label{sec_4.2.2_different Fint}

Vertical mixing shapes planetary atmospheres, especially when internal energy fluxes (\( F_{\rm int} \)) vary. Stronger vertical mixing heats deeper atmospheric layers when stellar and internal fluxes remain constant.
Figure~\ref{fig3}(A) shows how changes in \( F_{\rm int} \) affect temperature profiles and inhomogeneity. Solid lines represent no mixing, and dashed lines show mixing cases. Blue and green lines indicate inhomogeneous columns with \( F_{\rm int} = 0.001 \) and 0.005, respectively, while orange lines show a homogeneous column with \( F_{\rm int} = 0.003 \).

Without mixing, the internal temperature parameter \( K = 0.2 \). Upper atmosphere temperatures are the same across all cases, so the temperature ratio \( T_{\rm homo}^4/T_{\rm mean}^4 - 1 \), shown as a black line in Figure~\ref{fig3}(B), is zero. But when pressure exceeds \( P/P_{\rm s} > 0.3 \), the homogeneous column’s temperature matches that of a higher \( F_{\rm int} \) column. This raises \( T_{\rm homo}^4/T_{\rm mean}^4 - 1 \), peaking near \( P/P_{\rm s} \approx 1 \), as inhomogeneity grows.

\begin{figure*} 
    \centering
     \includegraphics[width=1.0\linewidth]{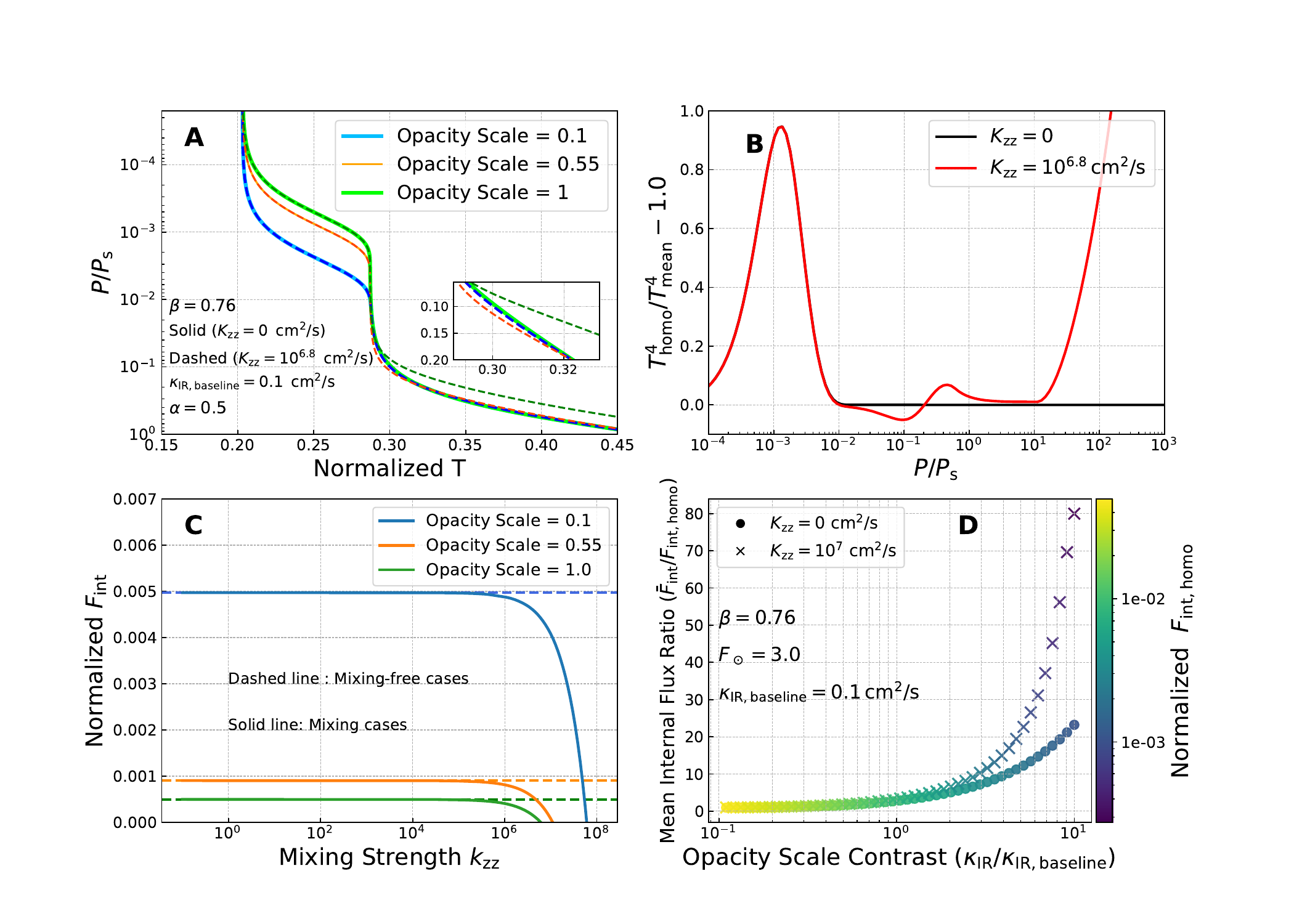}
    \caption{
   Panels (A)–(D) show vertical mixing effects in two inhomogeneous and one homogeneous RCME column of a giant planet’s atmosphere (\( \alpha = 0.5 \), \( \kappa_{\rm IR} = 0.1 \), 1.0 for inhomogeneous columns, 0.55 for homogeneous column, in \( \text{cm}^2/\text{g} \)).  
  \textbf{(A)} Temperature vs. pressure ratio (\( P/P_{\rm s} \)) profiles. Blue, green, and orange lines show \( \kappa_{\rm IR} = 0.1 \), 1.0, and 0.55, affecting temperature parameter \( K \). Solid lines: no mixing; dashed lines: mixing (\( K_{\rm zz} = 10^{6.8} \, \text{cm}^2/\text{s} \)).  
  \textbf{(B)} Temperature ratio (\( T_{\rm homo}^4 / T_{\rm mean}^4 - 1 \)) vs. pressure ratio (\( P/P_{\rm s} \)). Black lines: no mixing; red lines: mixing (\( K_{\rm zz} = 10^{6.8} \, \text{cm}^2/\text{s} \)).  
  \textbf{(C)} Internal temperature parameter (\( K \)) vs. mixing strength (\( K_{\rm zz} \)). Blue, green, and orange lines show \( \kappa_{\rm IR} = 0.1 \), 1.0, and 0.55. Solid lines: no mixing; dashed lines: mixing. 
\textbf{(D)} Mean internal flux ratio (\( \bar{F}_{\rm int}/F_{\rm int,homo} \)) at \( F_\odot = 3.0 \), baseline-model opacity \( \kappa_{\rm IR,baseline} = 0.1 \, \text{cm}^2/\text{g} \). Dots: no mixing; crosses: mixing (\( K_{\rm zz} = 10^{7} \, \text{cm}^2/\text{s} \)). Colorbar shows normalized \( F_{\rm int,homo} \).  
    }
    \label{fig4}
\end{figure*}

With vertical mixing, upper atmosphere temperatures remain uniform, matching no-mixing cases, as dashed lines in Figure~\ref{fig3}(A) and the red line in Figure~\ref{fig3}(B) confirm. But at \( P/P_{\rm s} \approx 0.2 \), mixing heats lower layers in inhomogeneous columns with lower \( F_{\rm int} \). Their \( K \) values rise sharply compared to no mixing, causing stronger bottom heating at lower pressures. Thus, \( T_{\rm homo}^4/T_{\rm mean}^4 - 1 \) in Figure~\ref{fig3}(B) increases earlier and exceeds no-mixing values. Inhomogeneity stabilizes when \( P/P_{\rm s} > 1 \).

Figure~\ref{fig3}(C) shows how \( K = T_{\rm s}^4 / \tau_{\rm s}^{\beta} \) changes with mixing strength. Blue and green lines represent inhomogeneous columns at \( F_{\odot} = 0.001 \) and 0.005, and the orange line shows a homogeneous column at \( F_{\odot} = 0.003 \). Below \( K_{\rm zz} \approx 10^6 \, \text{cm}^2/\text{s} \), \( K \) stays constant, with no effect on inhomogeneity. Above this threshold, \( K \) rises in low-\( F_{\rm int} \) columns, nearing the homogeneous column’s constant \( K \), indicating increased inhomogeneity. Higher \( K \) means stronger bottom heating. At high mixing strengths, \( K \) differences between columns shrink, showing greater heating for low-\( F_{\rm int} \) columns, as seen in Figure~\ref{fig3}(A)’s dashed lines.

\subsubsection{Effects of the Inhomogeneous IR opacity}\label{sec_4.2.3_different kappa}

The spatial and temporal opacity inhomogeneity in a planet's atmosphere, caused by horizontal material transport or chemical reactions, affects the internal temperature profile.
Figure~\ref{fig4}(A) shows temperature profiles vs.\ pressure ratio ($P/P_{\rm s}$) at fixed vertical mixing strength ($K_{\rm zz} = 10^{6.8}\,\mathrm{cm}^2/\mathrm{s}$), with baseline parameters $K_{\rm baseline} = 0.2$ and $\kappa_{\rm IR,baseline} = 0.1\,\mathrm{cm}^2/\mathrm{g}$. As opacity ($\kappa_{\rm IR}$) increases, the temperature parameter $K$ decreases ($K \propto \kappa_{\rm IR}^{-\beta}$), and normalized internal flux ($F_{\rm int}$) drops, as shown in Figure~\ref{fig1:}(B). Without mixing, the upper atmosphere has high temperature inhomogeneity: the orange line (homogeneous column, $\kappa_{\rm IR} = 0.55$) lies closer to the green line (inhomogeneous column, $\kappa_{\rm IR} = 1.0$) than the blue line ($\kappa_{\rm IR} = 0.1$). In deeper layers ($P/P_{\rm s} > 1$), inhomogeneity vanishes as all columns align.

Figure~\ref{fig4}(B) shows the temperature ratio ($T_{\rm homo}^4 / T_{\rm mean}^4 - 1$) vs.\ pressure ratio. Without mixing (black line), the ratio peaks at $P/P_{\rm s} \approx 0.002$, then drops to zero, indicating no inhomogeneity in deep layers. With mixing (red line), the ratio rises earlier ($P/P_{\rm s} \approx 0.01$), and then stabilizes at $P/P_{\rm s} \approx 0.5$, but increases again at $P/P_{\rm s} > 70$, showing stronger inhomogeneity at the base.

Vertical mixing enhances the greenhouse effect, heating the lower atmosphere and increasing inhomogeneity. Figure~\ref{fig4}(C) shows how mixing strength ($K_{\rm zz}$) affects $F_{\rm int}$ in inhomogeneous columns (blue: $\kappa_{\rm IR} = 0.1$; green: $\kappa_{\rm IR} = 1.0$) vs.\ a homogeneous column (orange: $\kappa_{\rm IR} = 0.55$). Without mixing, the homogeneous column's $F_{\rm int}$ (orange dashed line) is closer to the high-opacity column (green dashed line). 
Vertical mixing causes $F_{\rm int}$ to follow an exponentially decaying profile,  with the decline being more pronounced at lower opacities. 
Stronger mixing shifts the uniform column toward the high-opacity inhomogeneous column. 
Meanwhile, higher opacities amplify inhomogeneities, 
so that weaker mixing is sufficient to reduce $F_{\rm int}$.

As discussed in \S~\ref{sec_4.2.2_different Fint}, vertical mixing warms the lower atmosphere, counteracting the temperature decrease caused by the reduction in $F_{\rm int}$.  At higher opacities, the green column ($\kappa_{\rm IR} = 1.0$) shows stronger bottom heating, as seen in Figure~\ref{fig4}(A)'s inset. The homogeneous column (orange) aligns with the low-opacity column (blue) in deep layers. Thus, vertical mixing increases temperature inhomogeneity, especially at high opacities and pressures, driving significant heating in the planet's lower atmosphere.

Figure~\ref{fig4}(D) shows the internal flux ratio ($\bar{F}_{\rm int} / F_{\rm int,homo}$) between inhomogeneous and homogeneous columns vs.\ opacity contrast ($\kappa_{\rm IR} / \kappa_{\rm IR,baseline}$). The ratio, always above 1, rises sharply with increasing opacity contrast.
Higher opacity in inhomogeneous columns boosts the ratio. Vertical mixing amplifies this effect, driving a faster increase. At opacity contrast $>10$, the ratio is 80 times higher than the homogeneous case and 3.5 times higher than without mixing.

Vertical mixing enhances local cooling but reduces overall planetary cooling compared to no mixing. Crosses (mixing) in Figure~\ref{fig4}(D) show darker colors than dots (no mixing), showing a larger drop in $F_{\rm int,homo}$ at higher opacity contrast.
This drop slows global cooling.
Unlike Figure~\ref{fig2:}(D), where stellar flux contrast drives a smaller rise in $\bar{F}_{\rm int} / F_{\rm int,homo}$, opacity contrast has a stronger impact. Thus, opacity contrast significantly increases temperature structure inhomogeneity.

\section{Conclusion and discussion}\label{sec:conclusion}

Gas giant atmospheres show inhomogeneity due to uneven starlight, varying surface properties, and atmospheric processes. 
This inhomogeneity accelerates internal cooling in these planets.
Previous studies on radiative processes overlooked vertical mixing in the radiative layer of the atmosphere. This work investigates how gravity waves \citep{Lindzen1981,Strobel1987,Zhang_2020} and large-scale wave-driven vertical mixing \citep{1984Holton,Zhang_2020} impact atmospheric inhomogeneity and change the cooling efficiency within gas giants.

The cooling of gas giant interiors depends on the incoming stellar energy, opacity, and vertical mixing strength. With identical energy input and mixing intensity, vertical mixing considerably diminishes the response function $F_{\rm int}$, thereby reducing internal cooling. This effect is stronger when the visible opacity is low, and it increases with higher incoming energy. Even with a higher IR opacity, vertical mixing remains low $F_{\rm int}$.

This work uses the Jensen’s inequality to analyze temperature profiles in columns with homogeneous and inhomogeneous energy input. Inhomogeneous inputs lead to greater atmospheric inhomogeneity, especially in the upper layers. Vertical mixing enhances inhomogeneity. 
The increase in mixing intensifies the decrease in the response $F_{\rm int}$, increasing inhomogeneity.
As both the incident stellar flux contrast and the opacity contrast increase, the mean internal energy flux ratio $\bar{F}_{\rm int}/F_{\rm int,homo}$ rises, with the effect being more pronounced in cases of higher opacity contrast.
Thus, the cooling efficiency of a planet with vertical mixing increases at the local level. However, the total cooling is still lower than in the non-mixing case. In other words, vertical mixing ultimately suppresses the planet’s overall cooling. This is because both the $\bar{F}_{\rm int}$ of the two inhomogeneous columns and $F_{\rm int,homo}$ decrease, with the latter showing a more substantial reduction.

This work has certain limitations. Variability in incoming stellar radiation and internal opacity contributes to atmospheric inhomogeneity in gas giant planets. Opacity variations largely control the effectiveness of vertical mixing by influencing radiative and thermal gradients.  The semigray RCME radiative transfer model simplifies the study of the impact of vertical mixing on inhomogeneity by considering only visible and infrared opacity \citep{2010A&A...Guillot}. However, vertical mixing strength depends on density and opacity, and multi-band radiative transfer models, which account for hundreds of spectral bands with unique opacity coefficients \citep{Parmentier2015,2017MNRAS...Hubeny,Cavali2017Icar}, alter its effects. 
These variations significantly affect internal cooling rates. In addition, this study does not address how the chemical composition affects the opacity. Chemical components, sensitive to temperature changes, can significantly alter opacity, influencing the role of vertical mixing in planetary cooling.

To address these issues, we will improve our code to model realistic planetary opacity. Vertical mixing affects internal cooling and is linked to planetary habitability in the future. Studies show that three-dimensional planets naturally have uneven stellar radiation, surface, and atmospheric inhomogeneity. Future work will combine vertical mixing with habitability, exploring Mie scattering, chemical equilibrium or disequilibrium \citep{2017MNRAS...Hubeny}, atmospheric climate changes \citep{1988Kasting, 2023ApJ...Mukherjee}, shifts in the habitable zone \citep{2023Innes} and the lava planet \citep{2025Lava}. Including chemical components requires considering clouds \citep{2016AA...Fromang,2018MNRAS...Ryu,2018ApJ...Zhang,WallaceOrmel2019A&A} and hazes. We will also use new JWST data \citep{2024Wogan} to refine studies of the impact of vertical mixing on gas giant atmospheric inhomogeneity, ensuring that the results match the observations.

\begin{acknowledgments}
We thank the anonymous referee and  Xi Zhang for the valuable suggestions that significantly enhanced this work. 
This work has been supported by the Pre-Research Projects on Civil Aerospace Technologies of China National Space Administration (Grant No. D010303), the National Natural Science Foundation of China (Grant No. 42578010, T2495234, 4257080072), the Science and Technology Development Fund, Macau SAR (File No. 0139/2024/RIA2).
C.Y. has been supported by the National SKA Program of China (grant No. 2022SKA0120101), the National Natural Science Foundation of China (NSFC, No.12288102),  the science research grants from the China Manned Space Project (No. CMS-CSST-2021-B09, CMS-CSST-2021-B12, and CMS-CSST-2021-A10), International Centre of Supernovae, Yunnan Key Laboratory (No.202302AN360001), the National Natural Science Foundation of China (grants 11873103 and 12373071).
W.Z.  has been supported by the National Natural Science Foundation of China (Grant No.12503068), Funded by Basic Research Program of Jiangsu (Grant No.BK20251249), and Guangdong Basic and Applied Basic Research Foundation(Grant 2023A1515110805).
\end{acknowledgments}

\appendix
\section{Upward and Downward Fluxes for Generic RME Atmospheres}
\label{sec_appendix}

According to \citet{ZhangXi2023ApJ}, the upward and downward fluxes in the mixing scenario also can be expressed as follows
\begin{equation}
F^{+}(\tau)=\pi B\left(\tau_{\mathrm{s}}\right) e^{D\left(\tau-\tau_{\mathrm{s}}\right)}+2 \pi \int_\tau^{\tau_{\mathrm{s}}} B(t) e^{-D(t-\tau)} d t \ , 
\end{equation}
\begin{equation}
F^{-}(\tau)=-2 \pi \int_0^\tau B(t) e^{-D(\tau-t)} d t \ .
\end{equation}
Using RME temperature structure Equation (\ref{eq:sec4.2_Trad}), the 
downward flux $F^{-}(\tau)$ normalized by $4 \sigma T_0^4$ changes into 
\begin{equation}
\begin{aligned}
F_{\mathrm{rad}}^{-}(\tau)= & -\frac{1}{2} \int_0^\tau\left[S(t)+\left(D+D^2 t\right) F_{\mathrm{int}} + T_{\rm mix}^4 \right] e^{-D(\tau-t)} d t \\
= & -\frac{1}{2} \int_0^\tau\left[\frac{\partial F_{\mathrm{v}}}{\partial t}-D^2 \int_0^t F_{\mathrm{v}} d t-D F_{\mathrm{v}}(0) +\left(D+D^2 t\right) F_{\mathrm{int}} + T_{\rm mix}^4\right] e^{-D(\tau-t)} d t .
\end{aligned}
\label{eq:appendix_Frad_}
\end{equation}
In addition, 
\begin{equation}
\int\left(\frac{\partial F_{\mathrm{v}}}{\partial t}+D F_{\mathrm{v}}\right) e^{D t} d t=\int \frac{\partial\left(F_{\mathrm{v}} e^{D t}\right)}{\partial t} d t= F_{\mathrm{v}} e^{D t} \ .
\end{equation}
Equation (\ref{eq:appendix_Frad_}) will become
\begin{equation}
\begin{aligned}
F_{\mathrm{rad}}^{-}(\tau)= & -\frac{1}{2} \int_0^\tau\left[\left(\frac{\partial F_{\mathrm{v}}}{\partial t}+ D F_{\mathrm{v}}\right)- D\left(F_{\mathrm{v}}+ D \int F_{\mathrm{v}} d t \right) +D\left[\left.\left(D \int F_{\mathrm{v}} d t-F_{\mathrm{V}}\right)\right|_{\tau=0}+F_{\mathrm{int}}\right]+D^2 F_{\mathrm{int}} t + T_{\rm mix}^4  \right] e^{-D(\tau-t)} d t \\
= & -\frac{1}{2}\left[F_{\mathrm{v}}(\tau)-D \int_0^\tau F_{\mathrm{v}}(\tau) d \tau-F_{\mathrm{v}}(0)+D \tau F_{\mathrm{int}}\right]  -\frac{1}{2} \int_{0}^{\tau} T_{\rm mix}^4 e^{-D(\tau-t)}dt .
\end{aligned}
\end{equation}
Rearranging it gives
\begin{equation}
F_{\mathrm{rad}}^{-}(\tau)=-\frac{1}{2}\left[\frac{S(\tau)}{D}-\frac{1}{D} \frac{\partial F_{\mathrm{v}}}{\partial \tau}+F_{\mathrm{v}}(\tau)+D \tau F_{\mathrm{int}}\right] -\frac{1}{2} \int_{0}^{\tau} T_{\rm mix}^4 e^{-D(\tau-t)}dt \ .
\end{equation}
Given $F_{\rm rad}^{+} (\tau) +  F_{\rm rad}^{-} (\tau) = - F_{\rm v}(\tau)+ F_{\rm int}$ \citep{ZhangXi2023ApJ}, it is possible to derive the normalized $F_{\rm rad}^{+}$:
\begin{equation}
\begin{aligned}
    F_{\rm rad}^{+} (\tau) = & \frac{1}{2}\left[\frac{S(\tau)}{D}-\frac{1}{D} \frac{\partial F_{\mathrm{v}}}{\partial \tau}+F_{\mathrm{v}}(\tau)+D \tau F_{\mathrm{int}}\right] -\frac{1}{2} \int_{0}^{\tau} T_{\rm mix}^4 e^{-D(\tau-t)}dt - F_{\rm v}(\tau)+ F_{\rm int} \\
    = & \frac{1}{2}\left[\frac{S(\tau)}{D}-\frac{1}{D} \frac{\partial F_{\mathrm{v}}}{\partial \tau}-F_{\mathrm{v}}(\tau)+\left(2+D \tau \right) F_{\mathrm{int}}\right] -\frac{1}{2} \int_{0}^{\tau} T_{\rm mix}^4 e^{-D(\tau-t)}dt \ .
\end{aligned}
\label{eq:appendix_Frad}
\end{equation}

\bibliography{sample7}{}
\bibliographystyle{aasjournalv7}

\end{document}